\documentclass[runningheads]{llncs}
\usepackage{graphicx}

\begin{document}

\title{Assessing the Performance of Automated Prediction and Ranking of Patient Age from Chest X-rays Against Clinicians}

\titlerunning{Assessing the Prediction and Ranking of Patient Age from Chest X-rays}
\author{Matthew MacPherson\inst{1} 
\and Keerthini Muthuswamy\inst{2} 
\and Ashik Amlani\inst{2}
\and Charles Hutchinson\inst{1,3} 
\and Vicky Goh\inst{4,5} 
\and Giovanni Montana\inst{1,6}}


\authorrunning{M. MacPherson et al.}

\institute{University of Warwick, Coventry, UK, CV4 7AL \and
Guy's and St Thomas' NHS Foundation Trust, London, UK \and 
University Hospitals Coventry and Warwickshire NHS Trust, Coventry, UK \and
Department of Radiology, Guy's and St Thomas' NHS Trust, London, UK \and
School of Biomedical Engineering \& Imaging Sciences, King’s College London, London, UK \and
The Alan Turing Institute, London, UK
}

\maketitle

\begin{abstract}
Understanding the internal physiological changes accompanying the aging process is an important aspect of medical image interpretation, with the expected changes acting as a baseline when reporting abnormal findings. Deep learning has recently been demonstrated to allow the accurate estimation of patient age from chest X-rays, and shows potential as a health indicator and mortality predictor. In this paper we present a novel comparative study of the relative performance of radiologists versus state-of-the-art deep learning models on two tasks: (a) patient age estimation from a single chest X-ray, and (b) ranking of two time-separated images of the same patient by age. We train our models with a heterogeneous database of 1.8M chest X-rays with ground truth patient ages and investigate the limitations on model accuracy imposed by limited training data and image resolution, and demonstrate generalisation performance on public data. To explore the large performance gap between the models and humans on these age-prediction tasks compared with other radiological reporting tasks seen in the literature, we incorporate our age prediction model into a conditional Generative Adversarial Network (cGAN) allowing visualisation of the semantic features identified by the prediction model as significant to age prediction, comparing the identified features with those relied on by clinicians.

\end{abstract}

\section{Introduction}
The aging process is an important part of medical image analysis, since the physiological changes associated with age form a baseline for 'normality' in a given modality; observations which might be considered in the normal range for an elderly patient can indicate a pathological finding in a younger subject \cite{Hochhegger2012}. Understanding the characteristic features of the aging process as manifested in a particular modality is therefore of great importance in accurate and informative clinical reporting, and incorporating this knowledge into machine learning based diagnostic approaches could contribute to greater clinical utility than an 'age blind' model of abnormality presentation. Age prediction from chest X-rays is also known to be a challenging task for radiologists \cite{Gross1985}, and a greater understanding of the salient features learned in a reliable age prediction model could be of value in improving clinical understanding.

In this paper we leverage a set of 1.8M chest X-rays gathered from a range of clinical settings in six hospitals to train age prediction and ranking models for adult patients. We compare the performance of these models against three radiologists in a novel study testing age prediction and age ranking relative ability to explore clinicians' ability to perceive age-related changes in a single patient over time. Furthermore, to understand the features relied on by the models we use GAN-based age-conditional image generation, allowing progressive re-aging of a synthetic 'patient' and visualisation of age-relevant changes.

\subsubsection{Related Work.}
Deep learning has been used to estimate patient age in several modalities, including neurological MRI scans \cite{Cole2017} and paediatric hand X-rays \cite{Larson2018}. In the chest X-ray domain, regression models have been applied to the NIH ChestX-ray14 \cite{Wang2017} and CheXpert \cite{Irvin2019} public datasets, and report association of high prediction error with image abnormalities \cite{Karargyris2019,Sabottke2020}. Patient age and gender are predicted from a proprietary dataset with a CNN in \cite{Yang2021}, with Grad-CAM \cite{Selvaraju2017} used to visualise characteristic features. Recent work has shown that over-estimated patient age is predictive of cardiovascular and all-cause mortality rates \cite{Leki2021,Raghu2021}. GANs \cite{Goodfellow2014} have previously been used in a variety of medical image processing applications \cite{Yi2019}. Controllable image features in GAN images of mammograms, cell histology and brain MRI datasets have been explored recently\cite{Blanco21,Fetty2020,Ren2021}. While human face re-aging with GANs has previously been explored \cite{Huang2021}, to the best of our knowledge GAN-based age-conditional generation of chest X-rays has not been previously demonstrated.

\subsubsection{Contributions.} The main contributions of this work are: \textbf{(i)} We present a state-of-the-art chest X-ray age prediction model trained on a large heterogeneous dataset, with sensitivity analysis to training set size and image resolution, and show generalisation performance on the public dataset NIH ChestX-ray14. \textbf{(ii)} We present a study comparing the performance of human radiologists against our model on two tasks: (a) ground truth age prediction from a chest X-ray; (b) ranking two time-separated scans of the same patient in age order. \textbf{(iii)} We use GAN-generated synthetic chest X-rays conditioned on patient age to intuitively visualise age-relevant features via simulated age progression.

\section{Data and Methods}

\begin{figure}
    \includegraphics[clip, trim={0cm 0cm 0cm 0cm}, width=\linewidth]{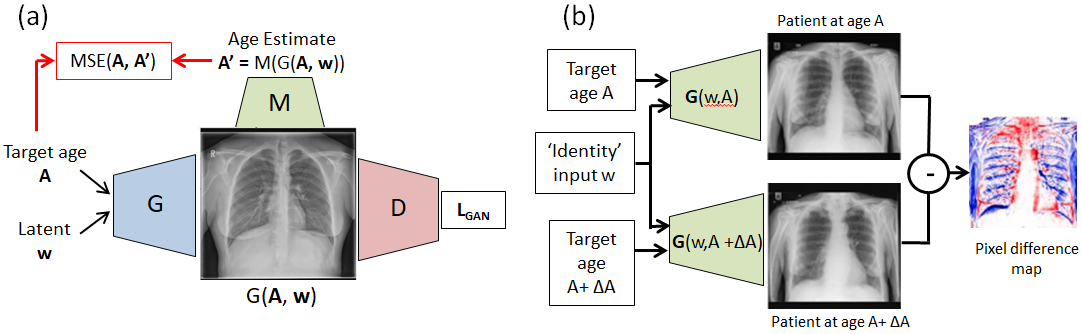}
    \centering
    \caption{(a) AC-GAN architecture for age-conditional image generation: G and D are the GAN Generator and Discriminator, and M is the pre-trained age prediction network (frozen during GAN training). (b) Aging feature visualisation via synthetic re-aging.}
\label{fig:workflow}
\end{figure}

\subsection{Dataset}
We train our models using a set of 1.798 million chest X-rays gathered under research partnerships with three Hospital Networks (UK NHS Trusts) comprising six hospitals covering a four million patient population: University Hospitals Coventry and Warwickshire NHS Trust, University Hospitals Birmingham NHS Foundation Trust, and University Hospitals Leicester NHS Trust. The data comprise scans taken from various departments and hardware representing both inpatient and outpatient settings covering the period 2006-2019, along with patient metadata including age, gender, anonymised patient ID, and free text radiologist's report. 
Images were normalised directly from the DICOM pixel data, and padded to square images with symmetric black borders where required.

After eliminating non-frontal and post-processed images, we obtained a core set of 1,660,060 scans from 955,142 unique patients, comprising 480,147 females and 474,995 males with mean age $61.2\pm19.0$ years. 688,568 patients had a single scan in the set, while 266,574 had multiple scans. The age distribution and longitudinal characteristics are presented in Fig.\ref{fig:dataset_plots}. The images represent an unfiltered, unaligned 'in-the-wild' set of images including both healthy patients and those with abnormalities, and to the best of our knowledge is the largest and most heterogeneous such dataset reported to date. We divide the images into a training set of 1,460,060 images, a validation set of 100k images, and a held-back test set of 100k images sampled from the same age and gender distribution.

\begin{figure}
    \includegraphics[clip, trim={0cm 0cm 0cm 0cm}, width=\linewidth]{paper1683_dataset_summary.png}
    \centering
    \caption{Characteristics of the dataset. (left) Patient age distribution stratified by gender. (right) Distribution of time separation between first and last images for multiple patient scans (70,493 patients separated by less than 3 months not shown).}
\label{fig:dataset_plots}
\end{figure}

\subsection{Age Prediction and Ranking Study} 
We performed a study aiming to address two questions: whether the ability of radiologists to order two images of the same patient by age is superior to their ability to estimate true patient age from a single image, and how their performance on each task compared to data-driven predictive models. To test this, we developed a web application which presented pairs of X-rays and required the users (working independently) to click on either the image they believed was older or a 'not sure' response, and also give an age estimate for one image (see supplementary material for the website interface). Our test data comprised 200 patients with pairs of images separated by up to 10 years, with 40 pairs from each two year separation bucket. The images were screened as 'normal' by a pre-trained NLP reading of the associated free-text radiologist's reports attached to the image \cite{Pesce2019}, ensuring that only physiological changes resulting from the aging process were present. By presenting pairs of time-separated images, we aimed to establish both the ability to estimate ground truth age (which previous work has indicated is a challenging task for which radiologists are not specifically trained), and over what time period age-related physiological changes become perceptible to an expert human observer with inter-subject differences removed. To the best of our knowledge this aspect has not been previously investigated.

\subsection{Age Prediction Models}
Given input images $x_{i}$ with continuous age labels $a_{i}$, we aim to train predictive models for patient age formulated as regression and ranking problems. We investigate the accuracy of alternate forms of output prediction model on a convolutional neural network (CNN) backbone: (a) linear regression, taking a single rectified output directly as an age estimate; (b) categorical classification  \cite{Rothe2018}, taking an output over nominal integer age labels $c_{j}, j\in[0,...,105]$ and treating the softmax-activated output as a probability distribution with the weighted sum $\sum {p_{j}c_{j}}$ as our estimate; and (c) ordinal regression \cite{Niu2016}, again outputting over integer age labels but applying sigmoid activation and treating each output as the probability of the age exceeding that label, estimating the age as $\sum{p_{j}}$. For pairwise image ranking we train a task-specific model, taking as input two images \{$x_{j}, x_{k}$\} with a binary label $y_{j,k}$ indicating if $a_{k} > a_{j}$. The images are passed through a shared CNN and the features concatenated before a fully connected module outputs a binary prediction of the ranking label. By learning simultaneously from a pair of images we expect the model to learn discriminative features directly relevant to fine-grained age discrimination and outperform on the ranking task.

For our CNN backbone we used the feature extraction layers of the EfficientNet-B3 architecture beginning from the ImageNet pre-trained weights. This was followed by the relevant fully connected prediction module with Xavier weight initialisation. An ADAM optimizer with initial learning rate 0.001, adaptive learning rate decay with a 0.5 decay factor and a three epoch patience level was used. We report test set results for the lowest validation set error epoch. All models were trained using an Nvidia DGX1 server with batch sizes 1024 ($224^{2}$ resolution), 512 ($299^{2}$ resolution), 224 ($512^{2}$ resolution) and 52 ($1024^{2}$ resolution).

\subsection{Age Conditional Image Generation and Manipulation}
To visualise the aging features identified by the model, we incorporated our best performing pre-trained age prediction model at $299^{2}$ resolution into a GAN to allow age-conditional synthetic image generation. By comparing images of the same synthetic patient at different ages with nuisance features such as pose held constant, we can visualise the features the age prediction model has identified as relevant to the aging process. To achieve this, we use an auxiliary classifier GAN (AC-GAN) architecture \cite{OdenaOlah2016}, illustrated in Fig.\ref{fig:workflow}, based on StyleGan2 \cite{Karras2020ada}. During training the generated images are passed into both the frozen pre-trained age prediction network $M$ to provide a training gradient for the age targeting accuracy of the generator $G$, and the discriminator network $D$ to enforce generated image realism. We concatenate the age target $A$ with the noise input ${w} \in \mathcal{R}^{512}$ into the StyleGan intermediate latent space. The network is trained over a GAN minmax value function with an additional MSE loss term between $A$ and the predicted age of the generated image $G(A, w)$:

\begin{equation}
\min _{G} \max _{D} V(D, G)= E_{{x}}[\log D({x})]+E_{{w}}[\log (D(G({A, w})))]  + \lambda(A - M(G(A, {w})))^{2}
\end{equation}

Leakage of age-relevant factors between ${w}$ and $A$ will degrade the age-prediction loss, therefore we expect aging features to be effectively separated from other 
factors of variation even though we do not explicitly enforce patient identity consistency. We therefore expect modifying $A$ with a fixed ${w}$ will perform synthetic re-aging of the same synthetic 'patient'. In order to visualise the age-related changes in the 'patient' we generate two images $\tilde{X}_{1} = G({A, w})$ and $\tilde{X}_{2} = G({A+\delta A, w})$, and generate a difference map of pixel intensity $\tilde{X}_{2} - \tilde{X}_{1}$ to visualise areas of greatest change in the image.

\section{Results}
\subsection{Age Prediction Accuracy}
\subsubsection{Model architectures \& resolution comparison.}
We aimed to establish the effect of different prediction models on age prediction accuracy, and summarise our results in Table \ref{tab1} (left) as mean absolute prediction error (MAE) in years. We observe little difference in accuracy between the three modelling approaches at $299^{2}$ resolution with MAEs of 3.33-3.34, leading us to surmise that the performance of the CNN is saturated with this size of dataset and the specific modeling approach is only marginally relevant. We also investigated using DenseNet-169 CNN (14.1M parameters) in place of EfficientNet-B3 (12M parameters), but obtained a slightly lower performance (3.40 vs 3.33 years). We elected to retain the simple regression model 'EfficientNet-LR' for the following work since a more complex approach does not appear to be supported by these results.

Increased image resolution improves model accuracy significantly up to $512^{2}$, with only a minor improvement seen from 3.05 to 2.95 MAE when increasing to $1024^{2}$ (Fig.\ref{fig:dataset_vs_mae}, right). We infer that age-relevant information is not purely limited to the coarse structural features in the image but is also present in intermediate-level features, while fine detail is less relevant. We observe that taking a mean of estimates at each of the four resolution levels for each image reduces MAE further to 2.78 years, which to the best of our knowledge is the best accuracy reported in a heterogeneous non-curated dataset. We note increased error with patient age, possibly attributable to higher levels of clinical abnormalities with age (see supplementary material).

\begin{table}
\caption{Summary MAEs of networks trained. Efficient=EfficientNet-B3, LR=Linear Regression, CL=Classifier, OR=Ordinal Regression.
}\label{tab1}
\centering
\begin{tabular}{|l|c|c|c|l|c|c|}
\hline
Network &  Resolution & MAE &   & Network &  Resolution & MAE \\
\hline
DenseNet+LR & 299x299 &  3.40 & & Efficient+LR &  224x224 & 3.64\\
Efficient+LR & 299x299 &  \textbf{3.33}  & & Efficient+LR &  299x299 & 3.33\\
Efficient+CL & 299x299 & 3.34 & & Efficient+LR & 512x512 & 3.05\\
Efficient+OR & 299x299 & 3.34 & & Efficient+LR & 1024x1024 & \textbf{2.95}\\
\hline
\end{tabular}
\end{table}

\subsubsection{Effect of Number of Training Images on Model Accuracy.}
\begin{figure}
    \includegraphics[clip, trim={0cm 0cm 0cm 0cm}, width=\linewidth]{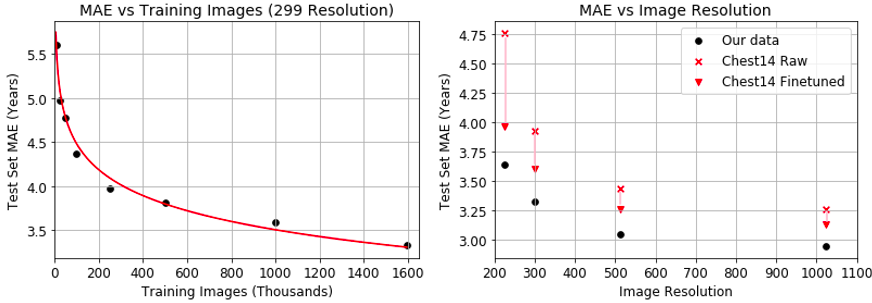}
    \centering
    \caption{(left) EfficientNet-LR model test set MAE ($299^{2}$ resolution) versus number of images in training data with logarithmic best fit. (right) Test set MAE versus training image resolution for our internal test set and Chest14.}
\label{fig:dataset_vs_mae}
\end{figure}

We investigated the extent to which the large dataset available contributes to prediction accuracy by training the Efficient-LR model at $299^{2}$ resolution on training sets from 10k to 1.597M images. Test set MAE is shown in Fig. \ref{fig:dataset_vs_mae} (left). MAE vs training set size empirically follows a logarithmic decline (shown as best fit), with diminishing improvements seen up to the full dataset size; we conclude that larger datasets are unlikely to be a major driver of improved performance.

\subsubsection{Generalisation Performance on NIH ChestX-ray14.} Performance of our pre-trained models on the Chest14 public dataset is shown in Fig. \ref{fig:dataset_vs_mae} (right). This dataset is provided as jpg files with unknown normalisation; we show results before and after fine-tuning on the Chest14 training data to adapt to the unknown normalisation. After fine-tuning we observe an MAE of 3.10 years at $1024^{2}$ resolution, ~5\% higher than for our own data; this could be due to information lost in pre-processing and image compression or different population characteristics and warrants further investigation. Direct performance comparisons with the literature are challenging, but we note that for the same dataset \cite{Leki2021} reports MAE of 3.78 years,  while \cite{Karargyris2019} reports 67.5\% within $\pm4$ years vs 87.3\% for our model. We conclude that pre-training on our dataset gives a significant performance increase over prior results.

\subsection{Clinician vs Algorithm Performance Study}
\begin{figure}
    \includegraphics[clip, trim={0cm 0cm 0cm 0cm}, width=\linewidth]{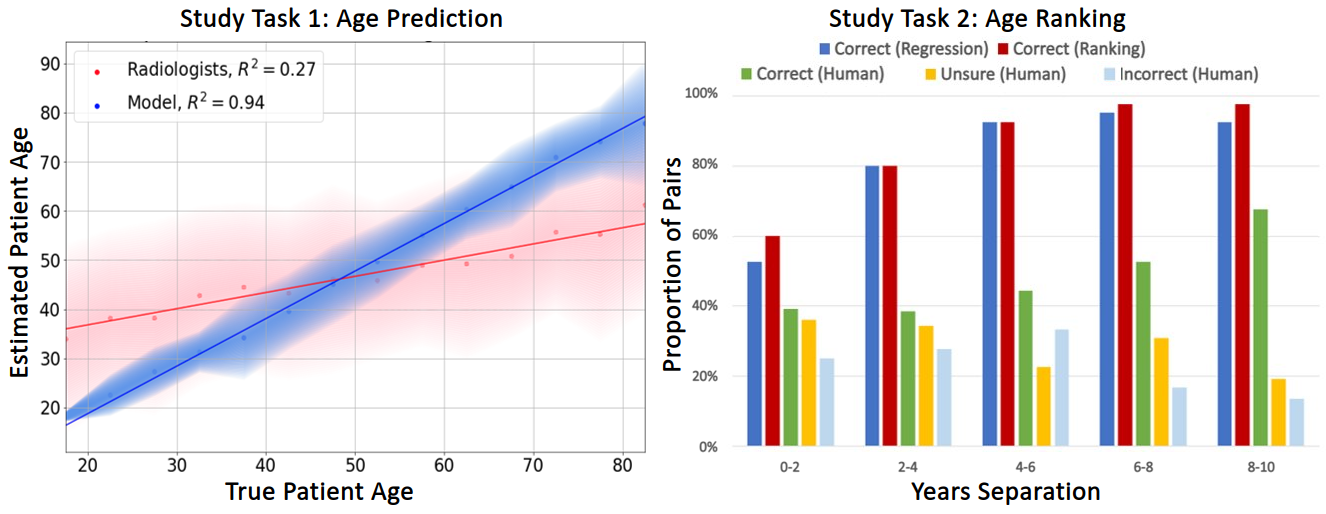}
    \centering
    \caption{Age prediction and ranking study results. (Left) Mean age prediction against true age bucketed in five-year intervals, with 2 SD areas shaded. (Right) Image ranking performance bucketed by age separation, regression \& ranking models vs humans.}
\label{fig:study_results}
\end{figure}

Results from the human vs model study are summarised in Fig.\ref{fig:study_results}. In the age estimation task (left) the regression model achieves an $R^{2}$ of 0.94, MAE 3.53 years and mean error (ME) $-2.23\pm 3.86$ years, whereas the radiologists achieve $R^{2}$ of 0.27 with MAE 11.84 years and ME $-3.12\pm14.25$ years. The accuracy of the three radiologists varied between MAE 10.86 and 12.69 years with $R^{2}$ 0.13 to 0.36. On the ranking task (right), both the ranking model and humans see increased accuracy with age separation as expected. The radiologists successfully ranked 48.3\% of all pairs and 67.1\% of those attempted, vs 82.5\% of all pairs for the regression model and 85.5\% for the ranking-specific model.

To test our hypothesis that humans are stronger at longitudinal change detection than point age estimation versus a null hypothesis of no benefit, we compare the observed ranking success with that expected if the two ages were estimated independently. Assuming Gaussian errors with the observed standard deviation of 14.25 years and calculating the probability of the sum of the two errors ($\sim\mathcal{N}$(0,(2x14.25)$^{0.5}$) exceeding the true age separation of each pair attempted, we derive an expected ranking success rate of $60.1\pm2.4\%$. Compared with the observed success rate of $67.1\%$ on attempted pairs, we conclude that the radiologists do benefit from the longitudinal information (p-value $0.002$). Performing the same analysis for the models with the observed point estimate standard deviation $3.86$ years, we expect a ranking success rate of $79.0\pm2.7\%$. We find (as expected) no evidence of the regression model benefiting from paired images (p-value 0.098), whereas we find the ranking model benefits from the paired information (p-value 0.008). 

\subsection{Aging Feature Visualisation}
\begin{figure}
    \includegraphics[clip, trim={0cm 0cm 0cm 0cm}, width=\linewidth]{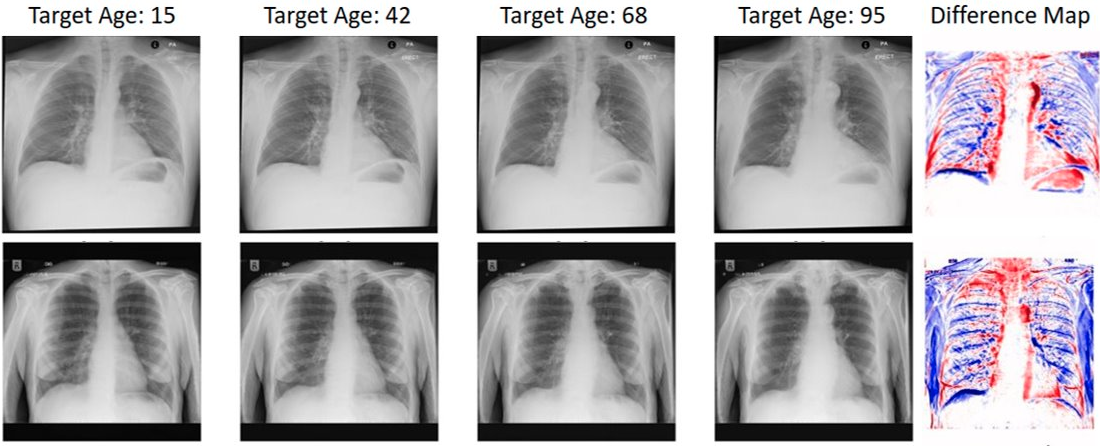}
    \centering
    \caption{Synthetic chest x-rays with varying target age and a constant patient 'identity' vector per row. The pixel difference between age 15 and 95 is shown in the final column highlighting the areas changed by the GAN to target the correct age.}
\label{fig:diff_maps}
\end{figure}

In Fig.\ref{fig:diff_maps} we show two examples of synthetic age interpolations. Each row uses a a constant ${w}$ 'patient identity' input, showing the same artificial patient at progressively older target ages. The pixel-wise difference map of the 'oldest' image versus the 'youngest' highlights the features changed by the GAN to manipulate the age. We observe a widening of the aortic arch, lowering and widening of the heart, apical shadowing, some narrowing of the rib-cage, and widening of the mediastinum as the major features relevant to age prediction. These features are consistent with the opinion of our consultant radiologists, although they note that decreased transradiancy and increased aortic calcification would be expected which are not evident in the synthetically aged images. Further examples are provided in the supplemental materials.

\section{Conclusion}
In this work we present a study comparing the performance of three radiologists on chest X-ray age prediction and ranking tasks, comparing with data-driven models trained on a highly heterogeneous non-curated set of chest X-rays from a variety of clinical settings in six hospitals. We conclude that (a) the radiologists are significantly more accurate at detecting age-related changes in a single patient than at estimating age in single images and (b) the models significantly outperform humans on both tasks. We report a state of the art MAE of 2.78 years on our proprietary test data, and demonstrate the highest reported generalisation performance on the public NIH Chest14 dataset. Our work indicates that accuracy gains are likely to be small from larger datasets, and that the majority of age-relevant information is present at $512^{2}$ resolution in this modality. We further demonstrate a GAN-based 'explainable AI' solution to visualise age-relevant features identified by the model, and compare with those identified by the radiologists based on their clinical experience.

\newpage

\newpage
\bibliographystyle{splncs04}
\bibliography{paper1683_mybibliography}

\begin{thebibliography}{10}
\providecommand{\url}[1]{\texttt{#1}}
\providecommand{\urlprefix}{URL }
\providecommand{\doi}[1]{https://doi.org/#1}

\bibitem{Gross1985}
BH, G., KF, G., KK, S., WM, W., FL, B.: Estimation of patient age based on
  plain chest radiographs. pp. 141--3 (6 1985)

\bibitem{Blanco21}
Blanco, R.F., Rosado, P., Vegas, E., Reverter, F.: Medical image editing in the
  latent space of generative adversarial networks. Intelligence-Based Medicine
  \textbf{5},  100040 (2021). \doi{10.1016/j.ibmed.2021.100040}

\bibitem{Cole2017}
Cole, J.H., Poudel, R.P., Tsagkrasoulis, D., Caan, M.W., Steves, C., Spector,
  T.D., Montana, G.: Predicting brain age with deep learning from raw imaging
  data results in a reliable and heritable biomarker. NeuroImage  \textbf{163},
   115--124 (12 2017). \doi{10.1016/j.neuroimage.2017.07.059}

\bibitem{Fetty2020}
Fetty, L., Bylund, M., Kuess, P., Heilemann, G., Nyholm, T., Georg, D.,
  Löfstedt, T.: Latent space manipulation for high-resolution medical image
  synthesis via the stylegan. Zeitschrift fur Medizinische Physik  \textbf{30},
   305--314 (11 2020). \doi{10.1016/j.zemedi.2020.05.001}

\bibitem{Goodfellow2014}
Goodfellow, I.J., Pouget-Abadie, J., Mirza, M., Xu, B., Warde-Farley, D.,
  Ozair, S., Courville, A., Bengio, Y.: Generative adversarial networks  (6
  2014), \url{http://arxiv.org/abs/1406.2661}

\bibitem{Hochhegger2012}
Hochhegger, B., Zanetti, G., Moreira, J.: The chest and aging: radiological
  findings. J Bras Pneumol  \textbf{38},  656--665 (2012),
  \url{https://www.researchgate.net/publication/233404468}

\bibitem{Huang2021}
Huang, Z., Chen, S., Zhang, J., Shan, H.: Pfa-gan: Progressive face aging with
  generative adversarial network. IEEE Transactions on Information Forensics
  and Security  \textbf{16},  2031--2045 (2021).
  \doi{10.1109/TIFS.2020.3047753}

\bibitem{Leki2021}
Ieki, H., et~al., K.I.: Deep learning-based chest x-ray age serves as a novel
  biomarker for 1 cardiovascular aging 2 3 . \doi{10.1101/2021.03.24.436773},
  \url{https://doi.org/10.1101/2021.03.24.436773}

\bibitem{Irvin2019}
Irvin, J., et~al., A.Y.N.: Chexpert: A large chest radiograph dataset with
  uncertainty labels and expert comparison p.~19, \url{www.aaai.org}

\bibitem{Karargyris2019}
Karargyris, A., Kashyap, S., Wu, J.T., Sharma, A., Moradi, M., Syeda-Mahmood,
  T.: Age prediction using a large chest x-ray dataset. p.~66. SPIE-Intl Soc
  Optical Eng (3 2019). \doi{10.1117/12.2512922}

\bibitem{Karras2020ada}
Karras, T., Aittala, M., Hellsten, J., Laine, S., Lehtinen, J., Aila, T.:
  Training generative adversarial networks with limited data. In: Proc. NeurIPS
  (2020), \url{https://github.com/NVlabs/stylegan2-ada-pytorch}

\bibitem{Larson2018}
Larson, D.B., Chen, M.C., Lungren, M.P., Halabi, S.S., Stence, N.V., Langlotz,
  C.P.: Performance of a deep-learning neural network model in assessing
  skeletal maturity on pediatric hand radiographs. Radiology  \textbf{287},
  313--322 (4 2018). \doi{10.1148/radiol.2017170236}

\bibitem{Niu2016}
Niu, Z., Zhou, M., Wang, L., Gao, X., Hua, G.: Ordinal regression with multiple
  output cnn for age estimation. vol. 2016-December, pp. 4920--4928. IEEE
  Computer Society (12 2016). \doi{10.1109/CVPR.2016.532}

\bibitem{OdenaOlah2016}
Odena, A., Olah, C., Shlens, J.: Conditional image synthesis with auxiliary
  classifier gans  (10 2016), \url{http://arxiv.org/abs/1610.09585}

\bibitem{Pesce2019}
Pesce, E., Withey, S.J., Ypsilantis, P.P., Bakewell, R., Goh, V., Montana, G.:
  Learning to detect chest radiographs containing pulmonary lesions using
  visual attention networks. Medical Image Analysis  \textbf{53},  26--38 (4
  2019). \doi{10.1016/j.media.2018.12.007}

\bibitem{Raghu2021}
Raghu, V.K., Weiss, J., Hoffmann, U., Aerts, H.J., Lu, M.T.: Deep learning to
  estimate biological age from chest radiographs. JACC: Cardiovascular Imaging
  \textbf{14}(11),  2226--2236 (2021).
  \doi{https://doi.org/10.1016/j.jcmg.2021.01.008},
  \url{https://www.sciencedirect.com/science/article/pii/S1936878X21000681}

\bibitem{Ren2021}
Ren, Z., Yu, S.X., Whitney, D.: Controllable medical image generation via
  generative adversarial networks. Electronic Imaging  \textbf{2021},
  112--1--112--6 (2 2021). \doi{10.2352/issn.2470-1173.2021.11.hvei-112}

\bibitem{Rothe2018}
Rothe, R., Timofte, R., Gool, L.V.: Deep expectation of real and apparent age
  from a single image without facial landmarks. International Journal of
  Computer Vision  \textbf{126},  144--157 (4 2018).
  \doi{10.1007/s11263-016-0940-3}

\bibitem{Sabottke2020}
Sabottke, C.F., Breaux, M.A., Spieler, B.M.: Estimation of age in unidentified
  patients via chest radiography using convolutional neural network regression
  (2020). \doi{10.1007/s10140-020-01782-5/Published},
  \url{https://pytorch.org/}

\bibitem{Selvaraju2017}
Selvaraju, R.R., Cogswell, M., Das, A., Vedantam, R., Parikh, D., Batra, D.:
  Grad-cam: Visual explanations from deep networks via gradient-based
  localization. vol. 2017-October, pp. 618--626. Institute of Electrical and
  Electronics Engineers Inc. (12 2017). \doi{10.1109/ICCV.2017.74}

\bibitem{Wang2017}
Wang, X., Peng, Y., Lu, L., Lu, Z., Bagheri, M., Summers, R.M.: Chestx-ray8:
  Hospital-scale chest x-ray database and benchmarks on weakly-supervised
  classification and localization of common thorax diseases. vol. 2017-January,
  pp. 3462--3471. Institute of Electrical and Electronics Engineers Inc. (11
  2017). \doi{10.1109/CVPR.2017.369}

\bibitem{Yang2021}
Yang, C.Y., Pan, Y.J., Chou, Y., Yang, C.J., Kao, C.C., Huang, K.C., Chang,
  J.S., Chen, H.C., Kuo, K.H.: Using deep neural networks for predicting age
  and sex in healthy adult chest radiographs. Journal of Clinical Medicine
  \textbf{10} (10 2021). \doi{10.3390/jcm10194431}

\bibitem{Yi2019}
Yi, X., Walia, E., Babyn, P.: Generative adversarial network in medical
  imaging: A review. Medical Image Analysis  \textbf{58} (12 2019).
  \doi{10.1016/j.media.2019.101552}

\end{thebibliography}

\newpage

\section{Supplementary Material}

\begin{figure}
    \includegraphics[clip, trim={0cm 0cm 0cm 0cm}, width=11cm]{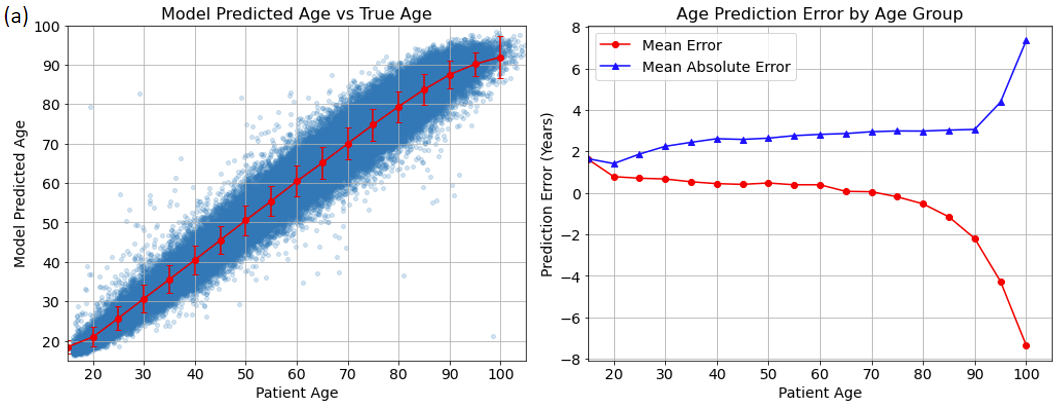}
    \centering
    \caption{ Efficient-LR ensemble model predictions vs ground truth for test set with 5-year bucketed mean and 1-sd overlay; MAE 2.78 years, $R^{2} = 0.963$. We note increased standard deviation with higher patient age, possibly reflecting increased variability of presentation of older patients with respect to abnormalities. Error increases sharply beyond 90 years.}
\label{fig:supplemental_model_errors}
\end{figure}

\begin{figure}
    \includegraphics[clip, trim={0cm 0cm 0cm 0cm}, width=10cm]{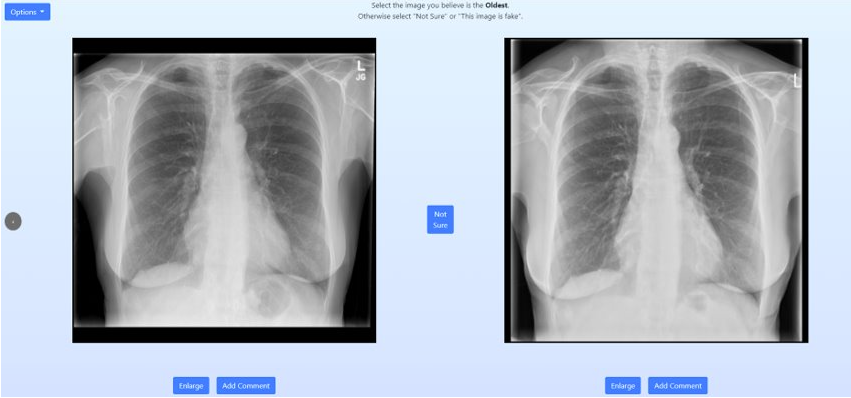}
    \centering
    \caption{Screenshot of age comparison web application used by the radiologists in our study. The user clicks on the image believed to show the patient at an older age or 'not sure' and enters an age estimate for one of the images before moving to the next pair.}
\label{fig:website_fig}
\end{figure}

\begin{figure}
    \includegraphics[clip, trim={0cm 0cm 0cm 0cm}, width=\linewidth]{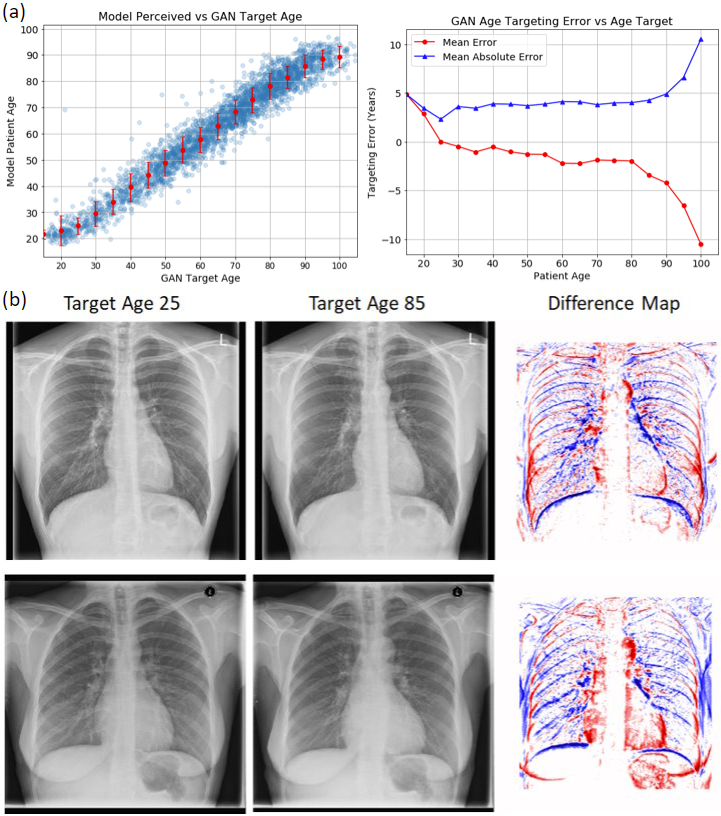}
    \centering
    \caption{(a) Target GAN image age vs model age estimate of generated image. The GAN struggles to generate images with perceived age $>90$ years. (b) Additional synthetic X-ray re-aging examples with a fixed identity vector per row. Unfolding of the aortic arch, bilateral widening and lowering of the heart outline, and narrowing of the ribcage are commonly identified features.}
\label{fig:supplemtal_diff_maps}
\end{figure}

\end{document}